# Microfluidic pressure-driven flow of a pair of deformable particles suspended in Newtonian and viscoelastic media: A numerical study

Giancarlo Esposito[1], Gaetano D'Avino[2], Massimiliano Maria Villone[2]

1. Laboratory of Fluid Mechanics and Rheology, Department of Chemical Engineering, University of Patras, Greece
2. Dipartimento di Ingegneria Chimica, dei Materiali e della Produzione Industriale, Università di Napoli Federico II, P.le Tecchio 80, 80125 Napoli, Italy

## Abstract

The manipulation and control of microparticles through non-intrusive methods is pivotal in biomedical applications such as cell sorting and cell focusing. Although several experimental and numerical studies have been dedicated to single suspended particles or clusters of rigid spheres, analogous cases with deformable particles have not been as thoroughly studied, especially when the suspending liquid exhibits relevant viscoelastic properties. With the goal of expanding the current knowledge concerning these systems, we perform a computational study on the hydrodynamic interactions between two neutrally buoyant initially spherical elastic particles suspended in Newtonian and shear-thinning viscoelastic matrices subjected to pressure-driven flow in a cylindrical microchannel. Due to the well-known focusing mechanism induced by both particle deformability and fluid elasticity, the two particles are assumed to flow at the axis of the tube. The rheological behavior of the viscoelastic continuous phase is modelled via the Giesekus constitutive equation, whereas the particles are assumed to behave as Neo-Hookean solids. The problem is tackled by employing a mixed finite element method. The effects of particle deformability, fluid elasticity, confinement ratio, and initial interparticle separation distance on the pair dynamics are investigated. The main outcome of this study is a quantitative indication of the flow conditions and spatial configurations (initial distances) under which the particles will spontaneously form organized structures. Such results are helpful to design efficient microfluidic devices with the aim of promoting particle ordering.

## 1. Introduction

The precise control of particle trajectories in microfluidic devices is crucial for several applications, e.g., sorting, separation, and encapsulation [1]–[4]. The possibility of exploiting "internal" forces, such as inertia or viscoelasticity, to drive the suspended particles towards specific positions within the cross-section of a channel has attracted significant interest in the recent years due to the simplicity and low cost of such techniques as compared to others. In particular, fluid elasticity is an effective mechanism to generate a single line of particles at the center of the cross-section of straight microchannels [5]–[7]. In extremely dilute systems, the aligned particles travel as isolated objects because the distances among them are so large that interparticle hydrodynamic interactions can be considered negligible. As the concentration of the particles increases, they get closer and start to hydrodynamically interact, leading to a variation of their separation distance while flowing through the channel. Recent works have shown that suspending fluid viscoelasticity promotes the formation of ordered microstructures [8, 9], i.e., the aligned particles tend to self-organize in a "train" with uniform spacing. The microstructure evolution depends on several parameters, such as the flow rate [8], fluid rheology [8, 10], confinement ratio [11], particle concentration [12] and size distribution [13, 14], wall slip [12], and geometric features of the channel [9]. The ordering induced by fluid viscoelasticity has been successfully applied to increase the efficiency of particle-in-droplet encapsulation [15, 16].

The behavior of a train is a complex phenomenon as it is governed by hydrodynamic interactions among multiple particles [17]. Nevertheless, many features of the microstructure evolution can be inferred by investigating the dynamics of a particle pair [8, 14] or triplet [17]. In the simplest case of two equal spherical rigid particles flowing at the centerline of a channel in a Newtonian fluid under inertialess conditions, numerical simulations have shown that the two particles do not change their relative position and travel at a velocity depending on the separation distance [18]. Fluid viscoelasticity strongly alters the pair dynamics, leading to different scenarios depending on the Deborah number (defined as the ratio between the fluid and the flow characteristic times), fluid rheology, confinement ratio, and initial particle separation [14, 19, 20]. Specifically, numerical simulations have shown that the particles attract or repel depending on whether their initial distance is below or above a critical value. The range of values of the initial interparticle distance for which the particles attract reduces as the Deborah number, the extent of the suspending fluid shear-thinning, and the confinement ratio increase [20]. At sufficiently high Deborah number and confinement ratio, the attractive dynamics disappears, thus the pair dynamics is repulsive whatever the interparticle distance. As experimentally verified by Del Giudice et al. [8], the repulsive dynamics promotes the formation of an ordered microstructure. The fact that the dynamics can be either attractive or repulsive has been attributed to the distribution of the axial component of the viscoelastic stress around the particles. At small distances, the fluid in the gap separating the two particles travels at approximately the same velocity of the pair, leading to a small velocity gradient that, in turn, produces low and uniform viscoelastic stresses. Hence, the axial stresses acting in front of the leading particle and behind the trailing one overcome those in the space between the two particles and push them closer to each other. On the contrary, as the distance between the two particles increase, high stresses appear in between the particles, leading to an inversion of the aforementioned scenario [8]. A similar study has been carried out on spheroidal particles with the major axis aligned with the channel centerline. Interestingly, as the spheroid aspect ratio increases, the range of values of the initial interparticle distance for which the particles attract reduces until it disappears regardless of the Deborah number and fluid shear-thinning. In this case, the large curvature near the tips of the spheroids leads to high stresses in between the particles also when they are close, thus promoting their repulsion [8].

More recently, the case of two spherical rigid particles with different size flowing in a viscoelastic fluid at the centerline of a square channel has been investigated by numerical simulations at finite Reynolds number [14]. Since the particle translational velocity reduces as the confinement increases, a string is always formed when the small particle is behind the large one. The particle chain formation is faster at high Reynolds and Weissenberg numbers and for more shear-thinning fluids.

When deformable particles are considered, their focusing at the center of the cross-section of a straight microchannel can be achieved also in inertialess Newtonian liquids due to the deformability-induced cross-streamline migration [21, 22]. On the other hand, very few works on the dynamics of a pair of deformable particles are available. Very recently, some numerical works based on the lattice Boltzmann method have dealt with the problem of the hydrodynamic interactions between two capsules while they are approaching the centerline of square- and rectangular-shaped channels filled with a Newtonian liquid under pressure-driven flow at non-negligible inertia, considering both equally-sized and non-equally-sized capsules [23]–[25]. However, none of those works investigates the dynamics of soft beads once they have reached the centerline of the channel. In addition, nothing is known on the dynamics of pairs of deformable particles at negligible inertia and in viscoelastic liquids.

In this work, we investigate the dynamics of a pair of initially spherical elastic beads in Newtonian and viscoelastic liquids under pressure-driven flow in a cylindrical tube. The particles are initially placed on the tube axis, thus assuming that the focusing induced by the intrinsic particle deformability and possibly by the suspending fluid elasticity has occurred upstream of the part of the channel under investigation [21]. The study is carried out by numerical simulations based on the finite element method (FEM) with an Arbitrary Lagrangian Eulerian (ALE) formulation. The viscoelastic liquid is modelled with the Giesekus constitutive equation, while the mechanical behavior of the particles is modelled with the neo-Hookean hyper-elastic constitutive equation. We perform a thorough analysis on the deformation of the particles and their distance and relative velocity as a function of the Deborah number, the elastic capillary number, the confinement ratio, and the initial inter-particle distance (the definition of such parameters being given below). The results of this work can provide a guide to the design of microfluidic devices promoting non-intrusive ordering of deformable particles, e.g., cells, and structure formation.

## 2. Mathematical formulation of the problem

### 2.1 Governing equations

The geometry of the computational domain is depicted in **Fig. 1**. Two non-Brownian initially spherical deformable particles with the same diameter $D_p$ lie on the axis of a cylindrical tube with diameter $D_c$ and length $L$ filled with a viscoelastic fluid subjected to Poiseuille flow. We denote by $d_0$ the initial distance between the surfaces of the particles and by $d$ their (time-dependent) actual distance. The initial position of the particles on the tube axis is justified by the fact that particle deformability and suspending fluid elasticity promote flow-focusing [21], thus it can be assumed that particles have aligned on the tube axis upstream of the part of the channel under investigation. Therefore, the geometry of the system has an axial symmetry, and we consider a cylindrical coordinate system with the $z$-axis oriented along the flow direction and the $r$-axis oriented along the tube radius.

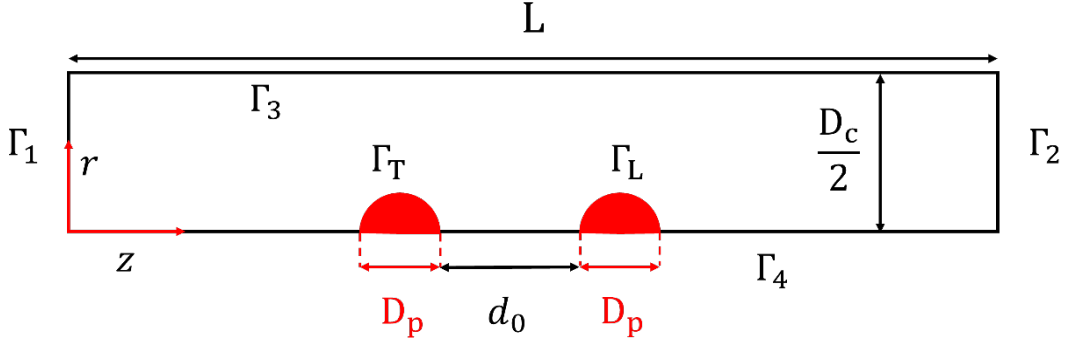

**Fig. 1**: Schematic representation of the system investigated in this work.

Under the assumptions that the system is isothermal, that both the suspending liquid and the suspended particles are incompressible, and that inertia and gravity are negligible, the governing equations for the system are the continuity equation (i.e., the mass balance) and the momentum balance equation, reading

$$\nabla \cdot \boldsymbol{u} = 0, \tag{1}$$

$$\nabla \cdot \boldsymbol{T} = \boldsymbol{0}, \tag{2}$$

where $\boldsymbol{u}$ is the velocity and $\boldsymbol{T}$ is the total stress tensor, which can be decomposed as follows into a hydrostatic component, depending on the pressure, and a deviatoric component, depending on the flow:

$$\boldsymbol{T} = -p\boldsymbol{I} + \boldsymbol{\sigma}, \tag{3}$$

where $p$ is the pressure, $\boldsymbol{I}$ is the identity tensor, and $\boldsymbol{\sigma}$ is the deviatoric part of $\boldsymbol{T}$, which can be further decomposed as

$$\boldsymbol{\sigma} = \eta_s \dot{\boldsymbol{\gamma}} + \boldsymbol{\tau}. \tag{4}$$

Here, $\eta_s$ is the "solvent" contribution to the dynamic viscosity, $\dot{\boldsymbol{\gamma}} = \nabla \boldsymbol{u} + \nabla \boldsymbol{u}^T$ is twice the rate-of-strain tensor, and $\boldsymbol{\tau}$ is the (visco)elastic stress tensor, for which a proper constitutive equation must be chosen. In this study, we model the viscoelastic continuous phase through the Giesekus model

$$\lambda \stackrel{\nabla}{\boldsymbol{\tau}} + \frac{\alpha \lambda}{\eta_p} \boldsymbol{\tau} \cdot \boldsymbol{\tau} + \boldsymbol{\tau} = \eta_p \dot{\boldsymbol{\gamma}}, \tag{5}$$

where the symbols $\lambda, \eta_p$, and $\alpha$ represent the relaxation time of the viscoelastic liquid, the "polymeric" contribution to the dynamic viscosity, and the mobility parameter, respectively. The physical meaning of $\alpha$ is associated with the anisotropy of the hydrodynamic drag and the Brownian motion experienced by polymeric molecules and it modulates shear-thinning. For $\alpha = 0$, the Giesekus constitutive equation degenerates into the constant-viscosity Oldroyd-B model, whereas, for $\lambda = 0$, it degenerates into the Newtonian constitutive equation with viscosity equal to $\eta_0 = \eta_s + \eta_p$. The choice of the Giesekus model is supported by its capability to adequately represent many viscoelastic effects. Indeed, it predicts non-zero first and second normal stress differences, along with a shear-thinning viscosity, and is commonly used to describe the behavior of polymeric solutions in microfluidics [26]. The symbol $\overset{\triangledown}{\boldsymbol{\tau}}$ identifies the upper-convected time derivative of $\boldsymbol{\tau}$, defined as

$$\overset{\triangledown}{\boldsymbol{\tau}} = \frac{\partial \boldsymbol{\tau}}{\partial t} + \boldsymbol{u} \cdot \nabla \boldsymbol{\tau} - (\nabla \boldsymbol{u}^T) \cdot \boldsymbol{\tau} - \boldsymbol{\tau} \cdot (\nabla \boldsymbol{u}). \tag{6}$$

The mechanical behavior of the particles is modelled by assuming $\eta_s = 0$ (i.e., purely elastic material) and by using the neo-Hookean hyperelastic constitutive equation for $\boldsymbol{\tau}$, which, in the velocity-based formulation, reads

$$\overset{\triangledown}{\boldsymbol{\tau}}_\mathrm{p} = G_\mathrm{p} \dot{\boldsymbol{\gamma}}, \tag{7}$$

with $G_\mathrm{p}$ the shear elastic modulus of the material constituting the particles. Such a model is commonly used to describe microgel beads and soft biological particles under moderate deformations [27]. In order to mathematically close the problem, we provide the following boundary conditions:

$$\boldsymbol{u}|_{\Gamma_1} = \boldsymbol{u}|_{\Gamma_2}, \tag{8}$$

$$(\boldsymbol{T} \cdot \boldsymbol{n})|_{\Gamma_1} = -(\boldsymbol{T} \cdot \boldsymbol{n})|_{\Gamma_2} - \Delta p \boldsymbol{n}, \tag{9}$$

$$\boldsymbol{\tau}|_{\Gamma_1} = \boldsymbol{\tau}|_{\Gamma_2}, \tag{10}$$

$$\boldsymbol{u}|_{\Gamma_3} = \boldsymbol{0}, \tag{11}$$

$$(\boldsymbol{n} \cdot \boldsymbol{u})|_{\Gamma_4} = 0, \tag{12}$$

$$(\boldsymbol{n} \cdot \boldsymbol{T} \cdot \boldsymbol{t})|_{\Gamma_4} = 0. \tag{13}$$

Equations (8) to (10) represent the periodicity conditions between the inlet and outlet boundaries for the velocity, the traction, and the viscoelastic extra-stress fields, with $\boldsymbol{n}$ the outwardly directed unit vector normal to the inlet section $\Gamma_1$ and the outlet section $\Gamma_2$, and the pressure drop $\Delta p$ in Eq. (9) to be computed. Equation (11) expresses the no-slip and no-penetration condition for the velocity of the liquid on the lateral wall of the cylindrical channel $\Gamma_3$. Eqs (12) and (13), with $\boldsymbol{t}$ the unit vector tangent to the symmetry axis $\Gamma_4$, enforce the symmetry condition on such boundary, nullifying the radial velocity and the shear stress. At the channel inlet, the flow rate $Q$ of the suspending medium is imposed as

$$Q = -\int_{\Gamma_1} \boldsymbol{u} \cdot \boldsymbol{n}\, dS. \tag{14}$$

At the interfaces between the carrier fluid (identified by the subscript "f") and the deformable particles (identified by the subscript "p"), we impose the continuity of the velocity and traction vectors

$$\boldsymbol{u}_f|_{\Gamma_{L,T}} = \boldsymbol{u}_p|_{\Gamma_{L,T}}, \tag{15}$$

$$(\boldsymbol{n} \cdot \boldsymbol{T}|_f)|_{\Gamma_{L,T}} = (\boldsymbol{n} \cdot \boldsymbol{T}|_p)|_{\Gamma_{L,T}}, \tag{16}$$

where $\Gamma_L$ identifies the interface between the leading particle and the suspending liquid and $\Gamma_T$ identifies the interface between the trailing particle and the suspending liquid.

The mathematical closure of the problem is obtained by imposing adequate initial conditions. Given the assumption of negligible inertia, those are required only on the (visco)elastic stress tensor in both phases. We adopt a stress-free initial condition for both the particles and the viscoelastic fluid, reading

$$\boldsymbol{\tau}|_{f,t=0} = \boldsymbol{\tau}|_{p,t=0} = \boldsymbol{0}. \tag{17}$$

The mathematical problem outlined above is made dimensionless by taking the diameter of the tube $D_c$ as the characteristic length, the average inlet velocity of the continuous phase $U_{av} = 4Q/(\pi D_c^2)$ as the characteristic velocity, the ratio $4Q/(\pi D_c^3 \eta_0)$ as the characteristic stress in the liquid phase (thus, we adopt a viscous scaling), and $G_p$ as the characteristic stress in the solid phase. Hence, the dimensionless variables (indicated with an overbar) are

$$\overline{\boldsymbol{\nabla}} = D_c \boldsymbol{\nabla}, \quad \overline{\boldsymbol{u}} = \frac{\pi D_c^2}{4Q} \boldsymbol{u}, \quad \overline{\dot{\boldsymbol{\gamma}}} = \frac{\pi D_c^3}{4Q} \dot{\boldsymbol{\gamma}}, \quad \bar{t} = \frac{4Q}{\pi D_c^3} t, \quad \overline{\boldsymbol{T}} = \frac{\pi D_c^3}{4Q\eta_0} \boldsymbol{T}, \quad \bar{p} = \frac{\pi D_c^3}{4Q\eta_0} p, \tag{18}$$

$$\overline{\boldsymbol{T}}_p = \frac{\boldsymbol{T}_p}{G_p},$$

Eqs. (1) to (7) can be rewritten in a dimensionless form as

$$\overline{\boldsymbol{\nabla}} \cdot \overline{\boldsymbol{u}} = 0, \tag{19}$$

$$\overline{\boldsymbol{\nabla}} \cdot \overline{\boldsymbol{T}} = \boldsymbol{0}, \tag{20}$$

$$\overline{\boldsymbol{T}} = -\bar{p}\boldsymbol{I} + \frac{\eta_s}{\eta_0} \overline{\dot{\boldsymbol{\gamma}}} + \overline{\boldsymbol{\tau}}, \tag{21}$$

$$\frac{4\lambda Q}{\pi D_c^3} \overset{\triangledown}{\overline{\boldsymbol{\tau}}} + \alpha \frac{4\lambda Q}{\pi D_c^3} \overline{\boldsymbol{\tau}} \cdot \overline{\boldsymbol{\tau}} + \overline{\boldsymbol{\tau}} = \overline{\dot{\boldsymbol{\gamma}}}, \tag{22}$$

$$\overset{\triangledown}{\bar{\boldsymbol{\tau}}}_\text{p} = \bar{\dot{\boldsymbol{\gamma}}}, \tag{23}$$

and the boundary condition expressed by Eq. (16) is reformulated as

$$\frac{4\eta_0 Q}{\pi D_\text{c}^3 G_\text{p}} \left[\boldsymbol{n} \cdot \left(-\bar{p}\boldsymbol{I} + \frac{\eta_\text{s}}{\eta_0} \bar{\dot{\boldsymbol{\gamma}}} + \bar{\boldsymbol{\tau}}\right)\right]\big|_{\Gamma_\text{L,T}} = \boldsymbol{n} \cdot (-\bar{p}\boldsymbol{I} + \bar{\boldsymbol{\tau}}_\text{p})\big|_{\Gamma_\text{L,T}}. \tag{24}$$

With this scaling, three dimensionless numbers appear in Eqs. (21), (22), (23) and (24). The first is the viscosity ratio $\xi = \eta_\text{s}/\eta_0$, modulating the relevance of the Newtonian solvent contribution to the total viscosity of the liquid. The second is the Deborah number $\text{De} = 4\lambda Q/(\pi D_\text{c}^3)$ and is given by the ratio of the characteristic times of the viscoelastic liquid (i.e., the relaxation time $\lambda$) and of the flow $t_\text{f} = \pi D_\text{c}^3/(4Q)$. As a Newtonian liquid has $\lambda = 0$, this is characterized by $\text{De} = 0$. The third dimensionless number is the elastic capillary number $\text{Ca}_\text{e} = 4\eta_0 Q/(\pi D_\text{c}^3 G_\text{p})$, given by the ratio between the viscous stresses exerted by the viscoelastic liquid on the particles and the elastic modulus of the particles. In addition, two geometrical dimensionless parameters are defined, namely, the confinement ratio $\beta = D_\text{p}/D_\text{c}$, representing the relative size of the particles with respect to the cross section of the tube, and the dimensionless distance between the particle surfaces $\bar{d} = d/D_\text{c}$. All the quantities appearing in the next sections are dimensionless, but they will be indicated without the overbar for brevity.

## 2.2 Numerical method and validation

To numerically solve the set of partial differential equations reported above, we employ a mixed finite element method (FEM) with linear interpolants for the stress and pressure fields and quadratic interpolants for the velocity field. The motion of the interface is tracked with an arbitrary Lagrangian Eulerian (ALE) formulation. To stabilize the convective term in the constitutive equation, we employ the SUPG technique [28] and the log-conformation approach is used to deal with the numerical difficulties arising in the treatment of the viscoelastic constitutive equation at moderate De [29]. The computational domain is discretized using triangular elements both in the continuous phase and in the particles. The elements at the interface between the solid and the liquid phases conform with the physical interface. A kinematic boundary condition is imposed at the interface between the two phases to guarantee the equality between the normal components of the mesh velocity and the physical velocity, whereas the tangential velocity of the nodes at the interface is such that the elements on the interface remain optimally distributed [30], thus the distortion of the mesh is strongly reduced. More details concerning the numerical method and the approach used to treat the solid-liquid interfaces as well as several validation cases in various geometries and flow conditions can be found elsewhere [31]. It is relevant to underline that our formulation is third-order accurate in space and second-order accurate in time. Since the only motion that the particles experience is along the axial direction, the mesh is moved rigidly along the flow direction with a velocity equal to the average velocity of the particle pair. In this way, we greatly reduce the distortion of the mesh elements. A typical grid used in this study is shown in **Fig. 2**. To ensure that our results are numerically independent of the grid resolution and time-step chosen, we conduct a mesh and time convergence procedure. To verify the adequacy of the grid resolution, we vary the number of elements on the particle-liquid interfaces $N_\text{p}$. Time convergence is checked by progressively reducing the time-step size. The relative velocity of the particles, i.e., the difference between their velocities $\Delta U_t = U_\text{L} - U_\text{T}$, where $U_\text{L}$ and $U_\text{T}$ are the velocities of the leading and the trailing particles computed in their centers of volume, respectively, is monitored as a function of time. **Fig. 3** shows mesh and time convergence results at $\beta = 0.4$, $d_0 = 0.1$, $\xi = 0.09$, $\text{De} = 1$, and $\text{Ca}_\text{e} = 0.1$. From the fair superposition of the data, we conclude that a mesh characterized by $N_\text{p} = 40$ (corresponding to a total number of triangular elements in the order of $10^4$) and a time-step $\Delta t = 7.5 \cdot 10^{-4}$ are adequate to obtain accurate results. (Just to make an example, the maximum relative discrepancy between the curves at $N_\text{p} = 30$ and $N_\text{p} = 40$ in **Fig. 3a** is of 2.57% and the time-averaged discrepancy is of 0.95%, whereas the maximum relative discrepancy between the curves at $N_\text{p} = 40$ and $N_\text{p} = 50$ is of 0.68% and the time-averaged discrepancy is of 0.54%.) These parameters satisfy mesh and time convergence even in the most critical cases, i.e., at the highest values of De and $\text{Ca}_\text{e}$. The minimum number of elements between the particles is fixed at $N_\text{p}/4$ in most of the cases, except when the particles are very close ($d_0 < 0.1$), where we linearly increase the number as the distance is reduced in order to guarantee higher accuracy.

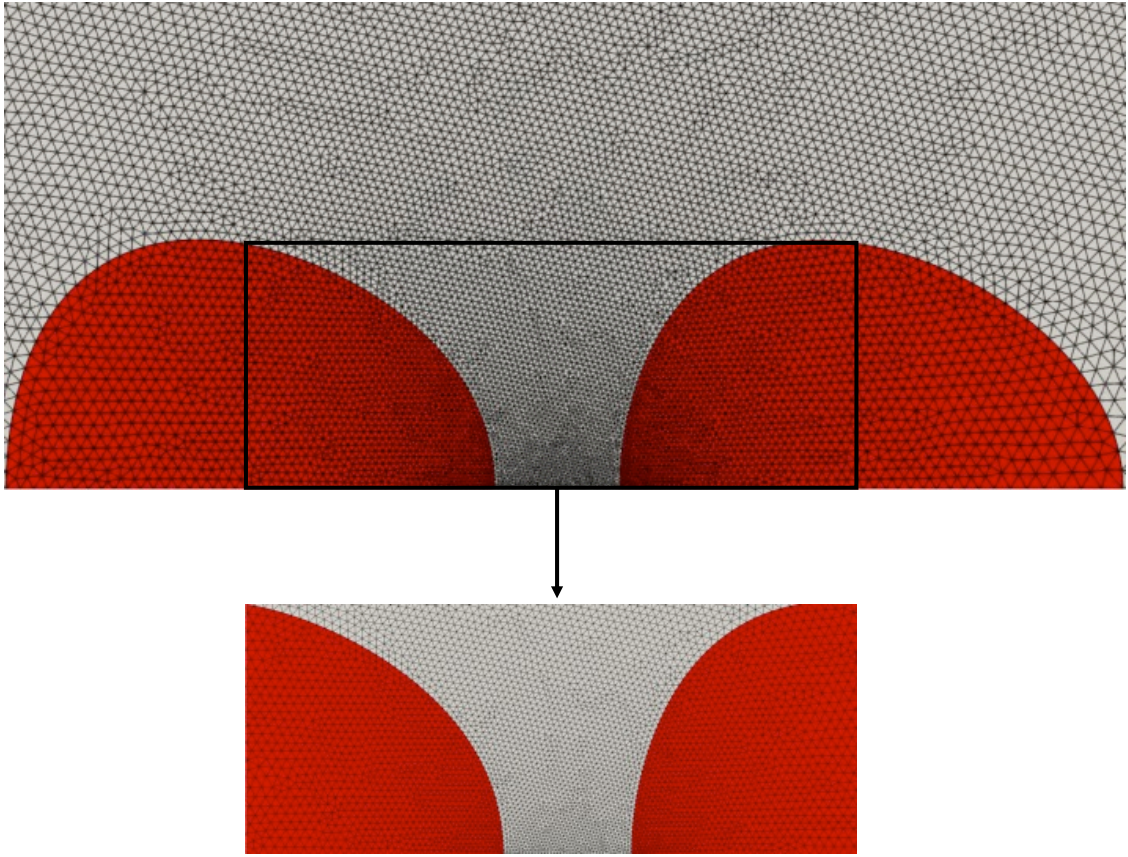

**Fig. 2**: Example of a mesh used in this study. The liquid domain is represented in white, whereas the particles are represented in red. The zoom at the bottom highlights the discretization of the portion of the domain in between the particles.

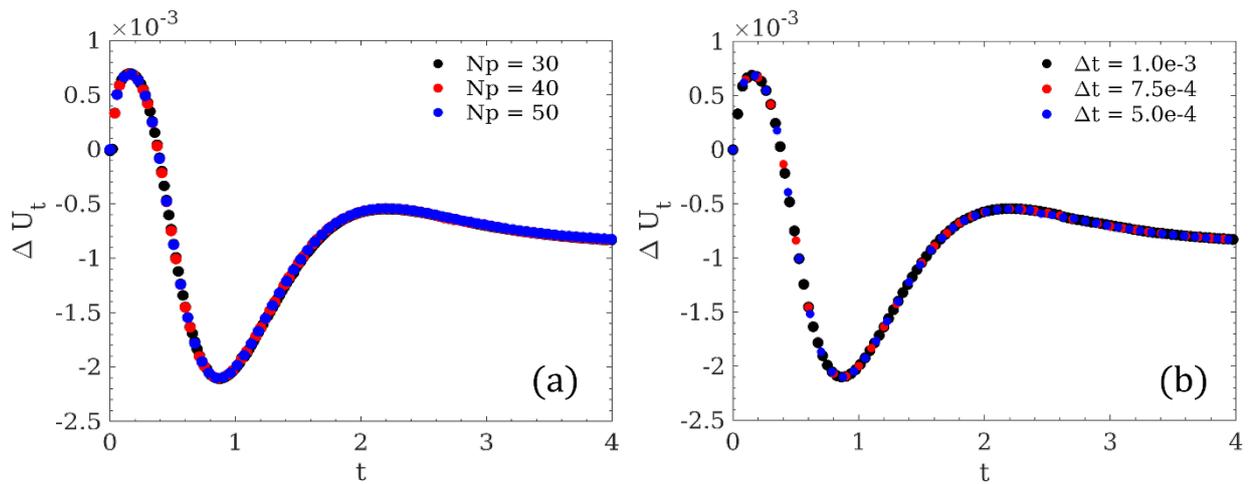

**Fig. 3**: Relative velocity of the particles as a function of time at given time-step and three different meshes (a) and at given mesh and three different time-steps (b). The parameters are $\beta = 0.4, d_0 = 0.1, \xi = 0.09, \text{De} = 1, \text{Ca}_e = 0.1$.

Finally, due to the periodic boundary conditions applied at the inlet and the outlet sections of the channel, we verify that the length of the domain $L$ is sufficient to avoid any interaction between the pair of particles and their periodic images along the flow direction. A total length $L = 50D_\text{p}$ is found to be adequate and consequently employed for all the calculations presented in this paper.

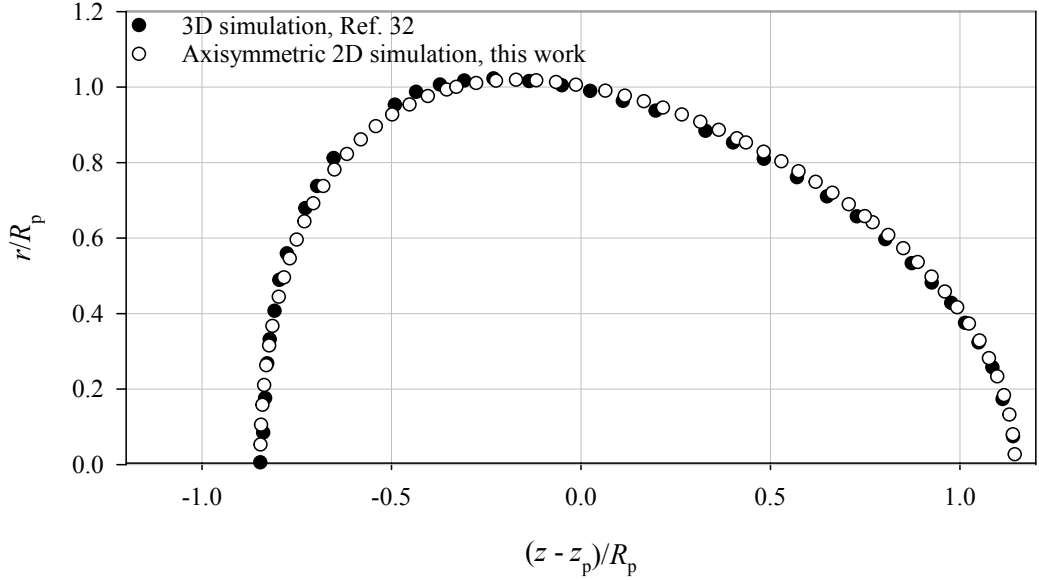

**Fig. 4:** Equilibrium deformed shape of a single particle suspended on the symmetry axis of a tube filled with a Newtonian fluid under pressure-driven flow at $\beta = 0.4$, $\xi = 0.09$, and $\text{Ca}_\text{e} = 0.1$. Filled circles: projection on the rz-plane of the shape obtained by Villone et al. [32] through a fully 3D simulation; empty circles: result of an axisymmetric 2D simulation.

As a validation of the code employed in this work, we perform the simulation of the flow-induced deformation of a single neo-Hookean particle suspended on the symmetry axis of a tube filled with a Newtonian fluid under pressure-driven flow at $\beta = 0.4$, $\xi = 0.09$, and $\text{Ca}_\text{e} = 0.1$ and we compare the equilibrium deformed shape of the particle with that obtained by Villone et al. [32] through a fully 3D finite-element simulation at the same values of the parameters, appearing in Fig. 2c of their work. The results of the two simulations are reported in **Fig. 4**, yielding an almost perfect overlap of the data.

## 3. Results

In this section, we present the results concerning the dynamics of a pair of deformable particles suspended in Newtonian and viscoelastic media. The first subsection is dedicated to the analysis of the hydrodynamic interactions in a Newtonian medium, exploring the effects of the deformability of the particles and their relative size with respect to the tube diameter. In the second subsection, a viscoelastic suspending medium is considered and consequently the influence of an additional parameter, namely, the Deborah number, is taken into account. The variables that allow us to quantitatively analyze the dynamics of the particle pair are the relative velocity of the particles $\Delta U_\text{t}$ and the Taylor deformation parameter $D$, which we define as the ratio between the difference and the sum of the particle maximum extension along the axial and the radial directions, as illustrated in **Fig. 5.** Due to the high number of dimensionless numbers governing the system, we fix some of them and perform a parametric study to assess the influence of the remaining parameters on the dynamics of the particle pair. Specifically, the viscosity ratio is set to $\xi = 0.09$ and, in case a viscoelastic suspending medium is considered, the mobility parameter is fixed to $\alpha = 0.2$.

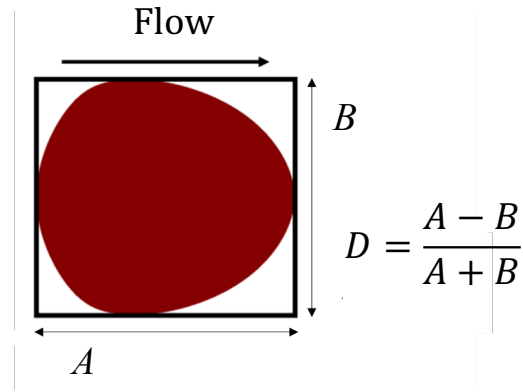

**Fig. 5:** Definition of the Taylor deformation parameter $D$.

### 3.1 **Pair of deformable particles in a Newtonian liquid**

It is well-known that a pair of rigid spherical particles suspended in an inertialess Newtonian fluid does not experience any relative motion due to the reversibility of the Stokes equations [18]. On the contrary, when the particles are deformable, the anisotropy induced by the flow-induced shape change leads to different translational velocities of the two particles, thus to a relative motion and a variation of their separation distance.

With the purpose of quantifying the effect of the deformability of the particles on their relative motion, we perform numerical simulations at a fixed confinement ratio and different values of the elastic capillary number.

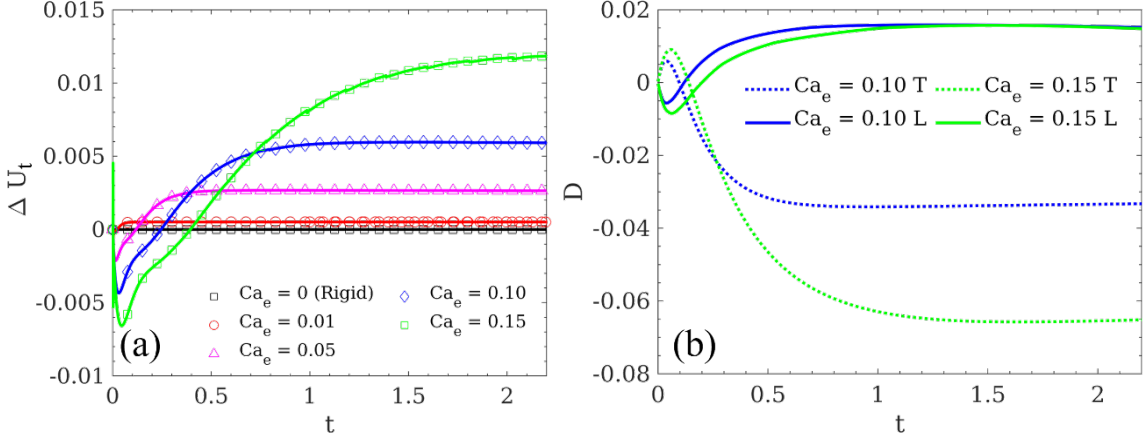

**Fig. 6:** (a) Relative velocity of the particles in a Newtonian medium as a function of time at $\beta = 0.4$, $d_0 = 0.1$, and five values of $Ca_e$ (see legend). (b) Deformation parameter of the leading (L) and the trailing (T) particle as a function of time at $\beta = 0.4$, $d_0 = 0.1$, and $Ca_e = 0.10, 0.15$.

**Fig. 6a** reports the evolution of the relative velocity of the particles at $\beta = 0.4$, $d_0 = 0.1$, and 5 values of $Ca_e$, as reported in the legend. The black line corresponds to the case where both particles are rigid ($Ca_e = 0$), showing that our simulations agree with the theoretical prediction of absence of relative motion between them. When a pair of deformable particles is considered, the situation qualitatively changes: after a phase characterized by a negative relative velocity (i.e., the particles get initially closer), a repulsive dynamics is observed with a steady-state terminal velocity that increases with the elastic capillary number. The duration of the transient phase monotonically increases with $Ca_e$ as the "inner" characteristic time is associated with the deformation of the particles.

In **Fig. 6b**, we report the dynamics of the deformation parameter of both the leading and the trailing particle at $\beta = 0.4$, $d_0 = 0.1$, and $Ca_e = 0.10, 0.15$. The particles are initially spherical, so $D$ is initially equal to 0. Comparing the results at $Ca_e = 0.10$ and $0.15$, it can be observed that the terminal $D$-values of the leading particle are substantially the same, whereas the magnitude of the deformation parameter of the trailing particle increases with $Ca_e$. It is noteworthy that the initial peak of the deformation parameter of the trailing particle is obtained at the same time as the minimum relative velocity, indicating that the particles initially tend to elongate, reducing their distance, and then slowly separate.

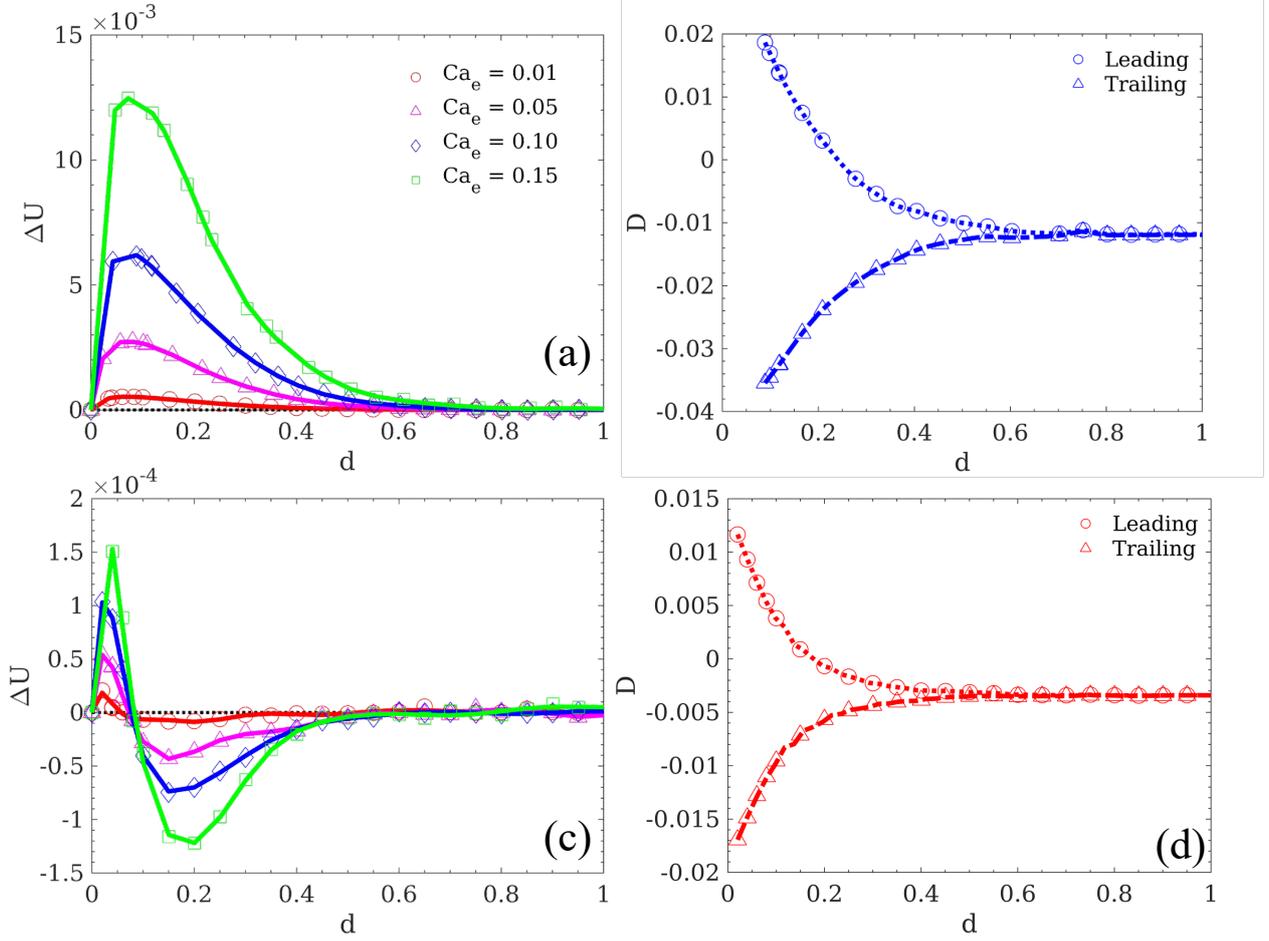

**Fig. 7**: Left: master curves of the terminal relative velocity of the particles in a Newtonian medium as a function of the interparticle distance $d$ at four values of $Ca_e$ (see legend) and $\beta = 0.4$ (a) and 0.2 (c). Right: master curves of the deformation parameter of the leading and the trailing particle as a function of the interparticle distance $d$ at $Ca_e = 0.1$ and $\beta = 0.4$ (b) and 0.2 (d).

Similarly to what has been observed in previous studies [19], we notice that, after a phase related to the development of the stresses inside the elastic particles, which is a consequence of the initial conditions on particle shape and stress, the terminal values of the relative velocity obtained at different values of $d_0$ and given values of the other parameters collapse on a master curve yielding the steady-state particle relative velocity $\Delta U$ as a function of the actual distance $d$. Similar master curves are found for the deformation parameter as well. The master curves of the terminal relative velocity at $\beta = 0.4$ and different values of the elastic capillary number are reported in **Fig. 7a**. Each marker on the curves corresponds to a simulation at a different value of $d_0$. The terminal relative velocity is always positive (i.e., the dynamics is always repulsive) and its magnitude increases with $Ca_e$. Above a distance $d \sim 0.6$, the relative velocity decreases asymptotically to 0 whatever $Ca_e$, namely, the particles do not interact anymore, thus behaving as if they were isolated. In **Fig. 7b**, the master curves of the Taylor deformation parameter are reported at $\beta = 0.4$ and $Ca_e = 0.10$, showing that the closer the particles the larger the magnitude of $D$ for both of them. Similarly to their relative velocity, the deformation of the particles is not influenced by their interaction above $d \sim 0.6$, thus both the leading and the trailing particle deform as isolated particles at those values of $\beta$ and $Ca_e$.

To elucidate the role of geometrical confinement on particle pair dynamics, we report in **Fig. 7c** and **Fig. 7d** the same quantities as in panels (a) and (b), but at $\beta = 0.2$. We find that, at lower confinement, the dynamics is still repulsive at short distances for any elastic capillary number, promoting the separation of the particles, yet, at larger distances, the particles tend to attract each other. Hence, differently from the more confined case at $\beta = 0.4$, a stable equilibrium distance is found at which the relative velocity of the particles is null, thus they keep such distance, neither attracting nor repelling. The effect of the particle deformability is mainly quantitative, since the magnitude of both the positive and the negative $\Delta U$-peaks increases with $Ca_e$. It is noteworthy that, although the intensity of both the repulsive and the attractive dynamics is strongly dependent on the deformability of the particles, the value of the stable equilibrium distance is almost independent of it. Furthermore, the dynamics of less confined particles is much slower than that of more confined ones (the magnitude of the particle relative velocity at $\beta = 0.2$ is almost 2 orders of magnitude lower than at $\beta = 0.4$).

Given $\beta$ and $d_0$, the particles undergo different deformation at different $Ca_e$. This can be justified by observing the distribution of the elastic stress inside them. **Fig. 8** displays the color maps of the axial component $\tau_{zz}$ of the elastic stress inside the particles at $\beta = 0.4$, $d_0 = 0.10$, and $Ca_e = 0.05, 0.15$. Although the two cases show qualitatively similar stress distributions, at $Ca_e = 0.15$, the inner core of the particles is characterized by a "more negative" (compressive) $\tau_{zz}$, while a "more positive" (extensional) $\tau_{zz}$ is found at the surface of the particles, promoting a higher deformation of the latters. Such observation also explains why the development of larger elastic stresses in softer particles enhances the repulsive dynamics: indeed, such stresses are transferred to the continuous phase, which "contrasts" the stresses acting at the interfaces of the two particles, resulting in the repulsion of the pair.

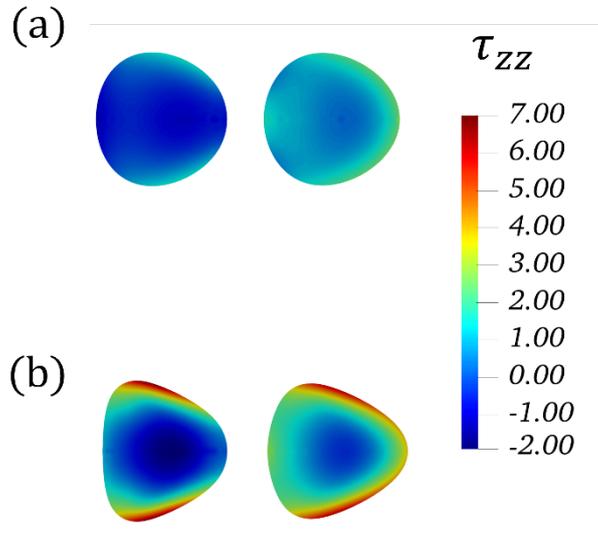

**Fig. 8:** Axial component of the elastic stress inside the particles at $\beta = 0.4$, $d_0 = 0.10$, and $\text{Ca}_e = 0.05$ (a) and 0.15 (b). The snapshots are taken after the initial viscoelastic stress development.

3.2    **Pairs of deformable particles in a viscoelastic liquid**

In this section, we analyze the behavior of a particle pair in a viscoelastic liquid. We carry out simulations at values of the blockage ratio between 0.2 and 0.6. We do not extend the analysis at larger $\beta$-values because it would not be technologically feasible to perform particle ordering in channels that are only slightly bigger than the particles due to clogging effects [11].

As in the case of a Newtonian matrix discussed above, we find the existence of master curves of the terminal relative velocity of the particles. In **Fig. 9,** we report them as a function of the interparticle distance $d$ at $\beta = 0.4$, four values of $\text{Ca}_e$ (as reported in the legend), and $\text{De} = 0.125$ (a), 0.25 (b), 0.5 (c), 1.0 (d), and 2.0 (e).

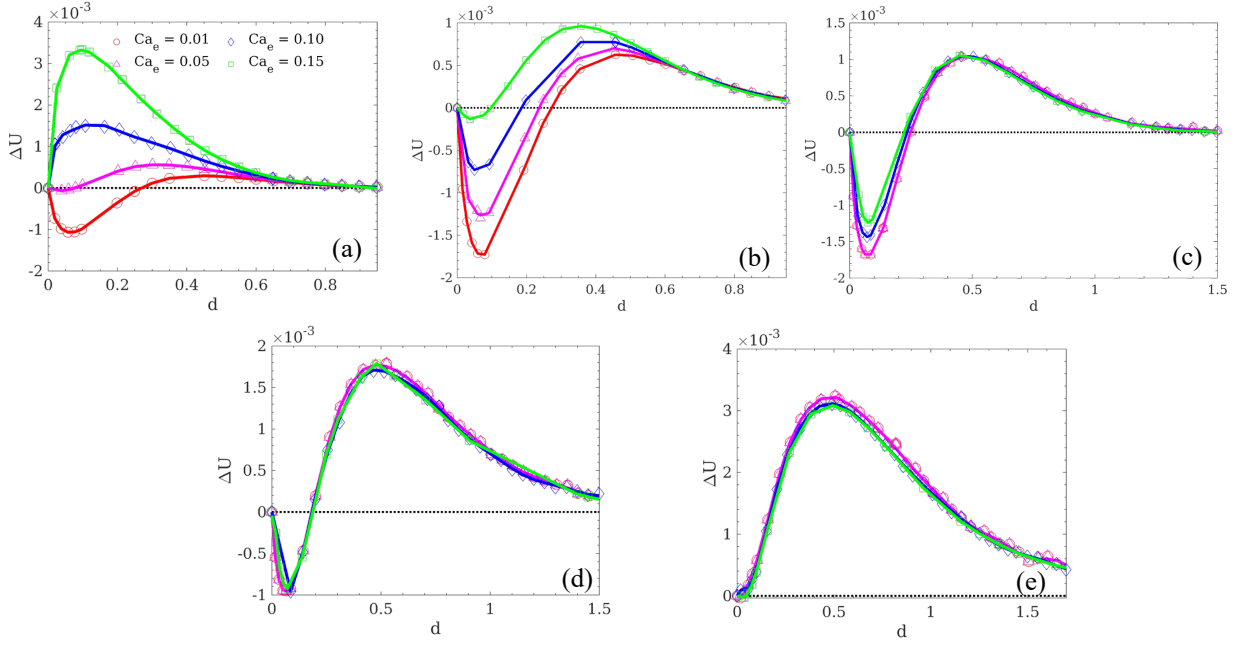

**Fig. 9:** Master curves of the terminal relative velocity of the particles in a viscoelastic medium as a function of the interparticle distance $d$ at $\beta = 0.4$, four values of $Ca_e$ (see legend), and De $= 0.125$ (a), $0.25$ (b), $0.5$ (c), $1.0$ (d), and $2.0$ (e).

The effect of the deformability of the particles on their interaction depends on the relevance of the elastic effects in the continuous phase, i.e., on the Deborah number. Indeed, when the elasticity is small, but finite, (De $= 0.125$, **Fig. 9a**) stiffer particles, having low $Ca_e$, show an attractive dynamics (below a critical $d$-value), whereas, when their deformability increases, the repulsion is promoted, first reducing the attractive region and, then, making it completely disappear. As the Deborah number increases to $0.25$ (**Fig. 9b**), a qualitatively similar behavior is observed at every $Ca_e$: at small interparticle distances, the dynamics is attractive ($\Delta U < 0$), whereas the particles repel each other at larger distances. Consequently, an unstable equilibrium distance exists, separating the two regions. Increasing $Ca_e$ has the effect of hindering the attractive dynamics, reducing the equilibrium distance. On the other hand, as $Ca_e \to 0$, we recover the asymptotic critical distance observed for rigid particles under similar conditions [19]. Whatever $Ca_e$, as the interparticle distance becomes much larger than the diameter of the tube, the relative velocity asymptotically decreases to zero and the particles behave like isolated ones. When the system is dominated by the elasticity of the continuous phase (i.e., when the Deborah number further increases), the deformability of the particles only has a small quantitative effect. As shown in **Fig. 9c**, **Fig. 9d**, and **Fig. 9e**, all the curves are nearly superimposed and the critical distance is progressively shifted at lower values as De increases, until it disappears at De $= 2$, meaning that the elastic response of the suspending medium promotes the repulsive dynamics, thus the ordering of the objects, as already proven for rigid particles [19]. To summarize, at $\beta = 0.4$, the deformability of the particles promotes their repulsion at low De, thus the particle ordering is more efficient with respect to the rigid case. When the elastic response of the suspending liquid is strong (i.e., high De), the attractive region shrinks until disappearing and the peak of the repulsive velocity increases, suggesting that the elasticity of the suspending medium is a key factor for the ordering mechanism.

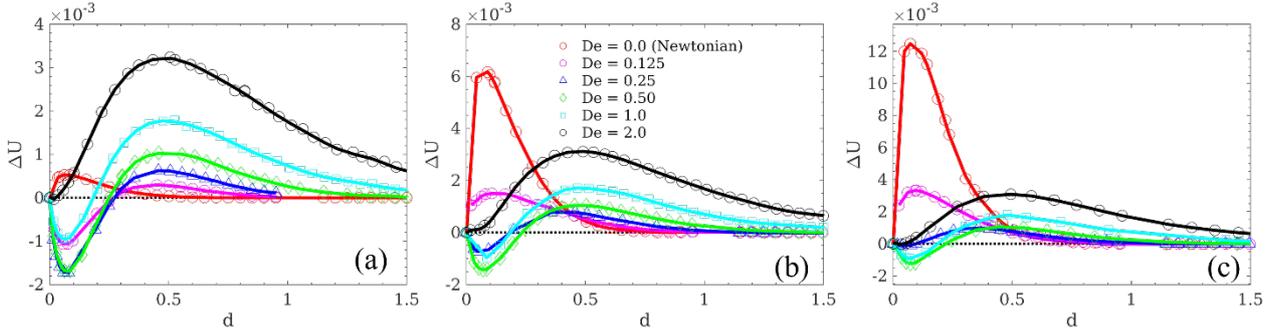

**Fig. 10:** Master curves of the terminal relative velocity of the particles in a viscoelastic medium as a function of the interparticle distance $d$ at $\beta = 0.4$, six values of De (see legend), and $Ca_e = 0.01$ (a), 0.1 (b), and 0.15 (c).

In **Fig. 10**, we display the master curves of $\Delta U$ as a function of $d$ at $\beta = 0.4$, 6 values of De (as reported in the legend), and $Ca_e = 0.01$ (a), 0.1 (b), and 0.15 (c). The results show that the effect of liquid elasticity is non-monotonic. When the suspending liquid is Newtonian (De = 0), the dynamics is always repulsive ($\Delta U > 0 \; \forall d$). For stiffer particles (low $Ca_e$, **Fig. 10a**), the increase of the Deborah number first makes the attractive region enlarge, then contract (until disappearing at De = 2.0), thus promoting a more efficient repulsive dynamics that is crucial to obtain particle ordering [8]. Stronger liquid elasticity leads to higher values of the peak of the relative velocity, thus allowing to achieve particle ordering in shorter tubes, which represents an important technological advantage. On the other hand, when the particles are softer ($Ca_e = 0.10$, **Fig. 10b**, and $Ca_e = 0.15$, **Fig. 10c**), we observe that the critical distance separating the attractive and repulsive regions first increases with the Deborah number (up to De = 0.5) and then reduces. The same holds true for the magnitude of the peak of the repulsive velocity.

The critical distance separating the attractive and repulsive regions is an unstable equilibrium point since, below that distance, the particles feel an attractive force that promotes their collision, whereas, above it, they tend to separate. Hence, although the relative velocity at the critical distance is null, any small perturbation of the interparticle distance would make the particles repel or attract. The trend of such a distance as a function of the elastic capillary number is illustrated in **Fig. 11**. Four curves are reported, corresponding to the cases at De = 0.0125, 0.25, 0.5 and 1.0. Particles characterized by $Ca_e \to 0$ show critical distances that are consistent with those obtained for rigid particles [17]. Increasing the deformability of the particles, thus, $Ca_e$, always reduces the critical distance $d_{cr}$, but its effect is more pronounced when the elastic response of the continuous phase is not relevant, i.e., at low De. On the contrary, $d_{cr}$ is almost independent of $Ca_e$ at De = 1.0.

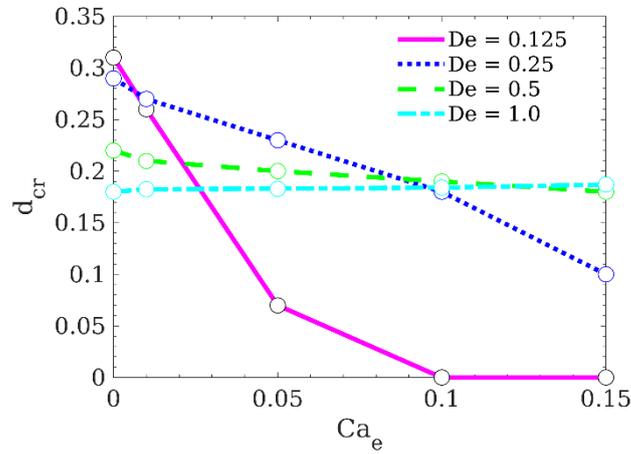

**Fig. 11:** Critical distance $d_{cr}$ as a function of the elastic capillary number at $\beta = 0.4$ and four values of the Deborah number (see legend).

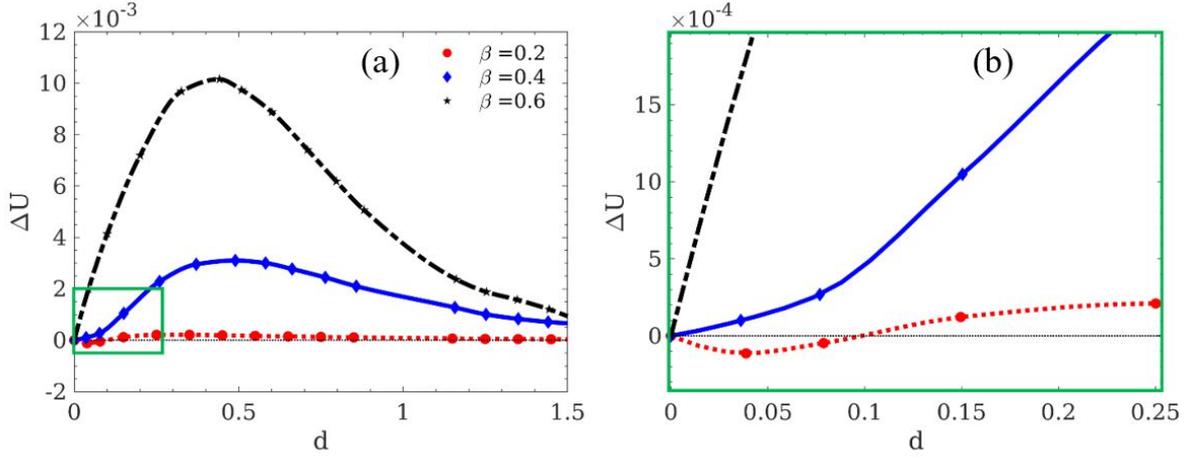

**Fig. 12:** (a) Master curves of the terminal relative velocity of the particles in a viscoelastic medium as a function of the interparticle distance $d$ at $Ca_e = 0.1$, $De = 2.0$, and three values of $\beta$ (see legend). (b) Zoom of the portion of panel (a) contained in the green rectangle.

In **Fig. 12a**, we display the master curves of the terminal relative velocity of the particles as a function of the interparticle distance $d$ at $Ca_e = 0.1$, $De = 2.0$, and $\beta = 0.2, 0.4$, and $0.6$, showing that the quantitative effect of the confinement ratio on the relative velocity is quite strong: a larger confinement speeds up the dynamics, with a peak of the repulsive velocity increasing by two orders of magnitude as $\beta$ goes from 0.2 to 0.6. Furthermore, attractive dynamics appears at low confinement ratio even at the largest investigated De (see **Fig. 12b** reporting a zoom of the portion of panel (a) contained in the green rectangle). The effect of the deformability of the particles is reversed at high confinement ratios, as it appears by comparing the two panels of **Fig. 13**, which shows the master curves of the terminal relative velocity of the particles as a function of $d$ at $De = 1.0$, different values of $Ca_e$ (as indicated by the legend), and $\beta = 0.2$ (a) and 0.6 (b). Such observation can be explained again by the analysis of the perturbed stress field around the particles, displayed in **Fig. 14**, where it emerges that the magnitude of the axial viscoelastic stresses around the particles is larger in the most confined case, promoting particle deformation and increasing the relative velocity. Analogous comments can be made on the perturbed pressure field, displayed in **Fig. 15.**

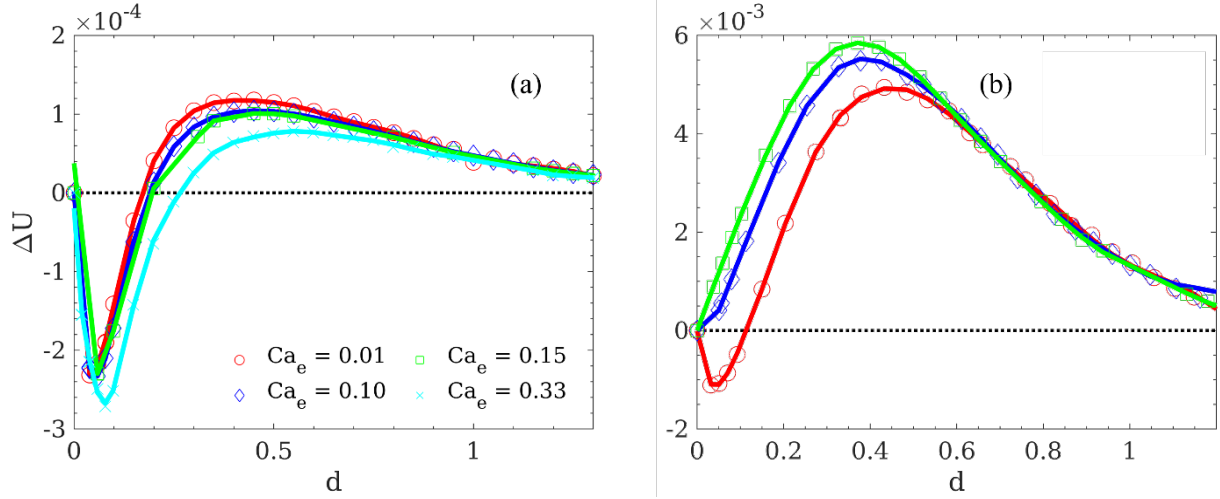

**Fig. 13**: Master curves of the terminal relative velocity of the particles in a viscoelastic medium as a function of the interparticle distance $d$ at De $= 1.0$, different values of $Ca_e$ (see legend), and $\beta = 0.2$ (a) and 0.6 (b).

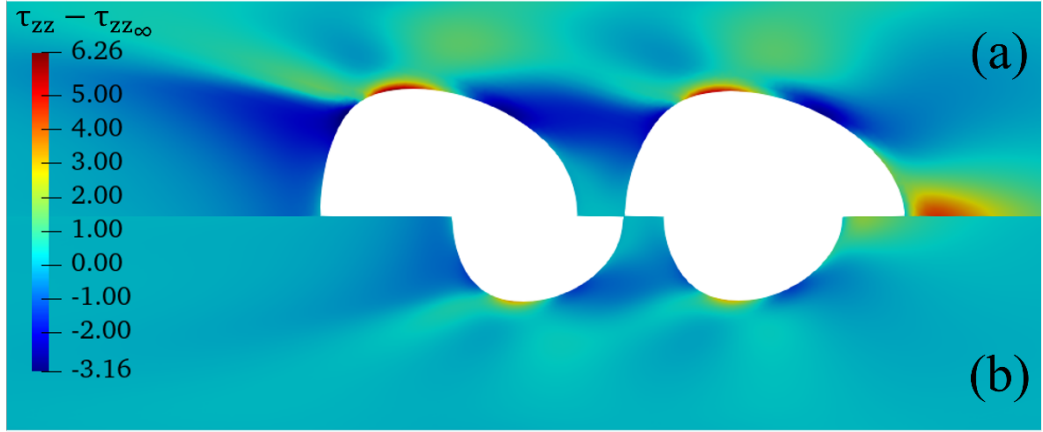

**Fig. 14:** Perturbation of the axial viscoelastic stress field in the liquid around the particles at $d_0 = 0.10$, $Ca_e = 0.15$, De $= 0.5$, and $\beta = 0.6$ (a) and 0.4 (b). $\tau_{zz_\infty}$ is the axial viscoelastic stress field in the liquid in the absence of particles. The snapshots are taken after the initial viscoelastic stress development.

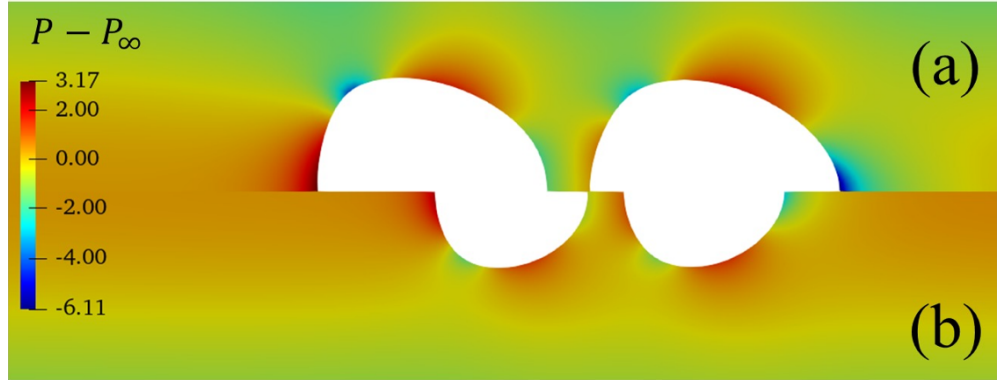

**Fig. 15:** Perturbation of the pressure field in the liquid around the particles at $d_0 = 0.10$, $Ca_e = 0.15$, $De = 0.5$, and $\beta = 0.6$ (a) and 0.4 (b). $P_\infty$ is pressure in the liquid in the absence of particles. The snapshots are taken after the initial viscoelastic stress development.

As previously reported for single elastic particles in viscoelastic liquids [32], the elasticity of the suspending medium has a quantitative effect on the deformation of the particles. In particular, more elastic fluids hinder the deformation. **Fig. 16** displays the shapes of the particles at $\beta = 0.4$, $d_0 = 0.10$, and $Ca_e = 0.15$ in the Newtonian case (red) and at increasing values of the Deborah number. The particle shapes are progressively less deformed as De increases. On the other hand, as just discussed, the deformation is promoted by geometrical confinement due to the stronger particle-wall hydrodynamic interactions leading to higher stresses in the gaps between the particles and the wall. **Fig. 17** reports the shapes of the particles at $d_0 = 0.10$, $Ca_e = 0.15$, $De = 0.5$, and different values of $\beta$, highlighting the change from the nearly-circular to the bullet-like shape as the confinement increases. It is worth remarking that, in all the cases, the particles deform differently, with a more pronounced deformation of the leading particle. This is responsible for the non-zero relative velocity of the two particles.

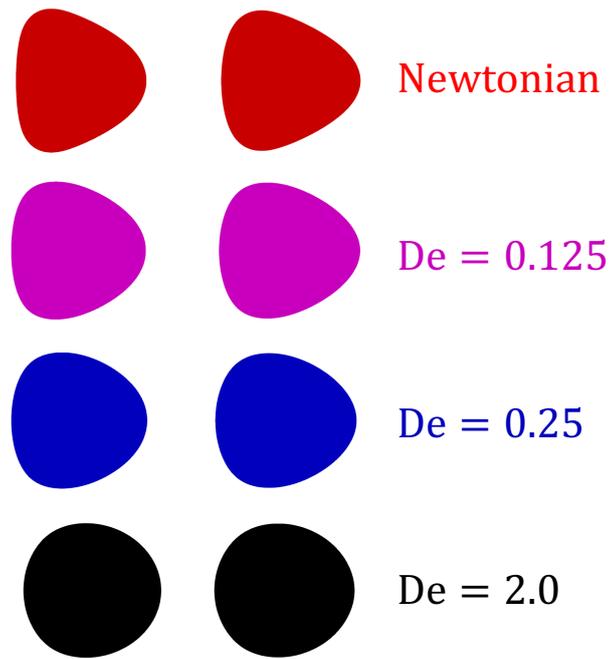

**Fig. 16:** Comparison of the steady state shapes of the particles at $\beta = 0.4$, $d_0 = 0.10$, $Ca_e = 0.15$, and different values of De.

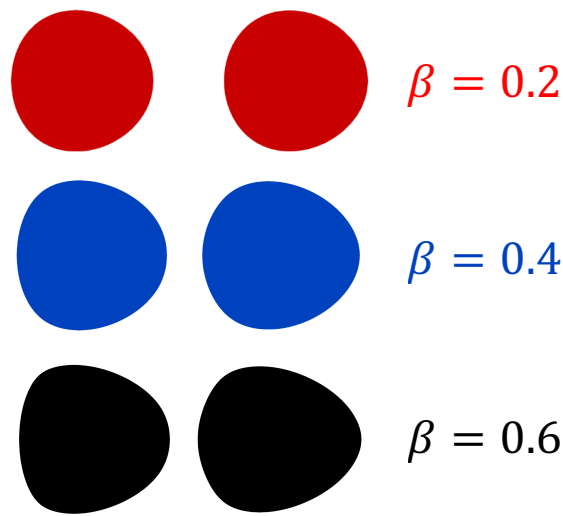

**Fig. 17:** Comparison of the steady state shapes of the particles at $d_0 = 0.10$, $Ca_e = 0.15$, De = 0.5, and different values of $\beta$.

In summary, the rheology of the suspending fluid strongly affects the dynamics of a pair of deformable particles. In a Newtonian liquid, depending on the confinement ratio, the pair either reaches a stable equilibrium distance or the two particles repel for any separation distance and the relative velocity is higher as the deformability of the particles increases. In a viscoelastic liquid, different scenarios arise depending on the strength of the elastic response of the suspending medium. When the particles are suspended in a mildly elastic liquid and the confinement is significant, the deformability of the particles promotes a repulsive dynamic, which allows their ordering. On the other hand, when the elasticity of the continuous phase is dominant, the deformability of the particles does not play a crucial role and the pair dynamics is qualitatively similar to that of two rigid particles, with the existence of a critical distance below which the particles approach and above which they separate. Consequently, the deformability of the particles represents a suitable feature to promote their ordering in Newtonian and mildly elastic liquids.

## 4. Conclusions

We carried out arbitrary Lagrangian Eulerian finite-element numerical simulations to explore the hydrodynamic interactions of a pair of equal initially spherical non-Brownian deformable particles suspended in Newtonian and viscoelastic media subjected to Poiseuille flow in a cylindrical tube. The particles are initially placed along the axis of the tube, thus we assume that the focusing mechanism has already occurred, and their relative distance, velocity, and deformation are tracked.

In a Newtonian liquid at relatively high confinement ratio, the deformability of the particles produces a repulsive effect until their distance becomes sufficiently large so that the particles behave as isolated objects and their relative distance remains fixed in time. As the confinement ratio decreases, attractive dynamics is observed beyond a critical distance. Hence, a stable equilibrium separation distance appears. The analysis is then extended to a viscoelastic suspending fluid, showing that the viscoelasticity of the fluid and the deformability of the particles act synergically to promote the repulsion of the particles when the confinement ratio is high, thus promoting particle ordering. When the deformability of the particles is weak (i.e., at very small elastic capillary number), our results are in quantitative agreement with previous studies concerning rigid spheres.

The findings presented in this work are useful to design microfluidic devices to promote non-intrusive ordering of deformable particles, e.g., cells, and structure formation. We plan to extend this study towards the analysis of triplets and trains of deformable particles and also to consider different rheological behaviors of the suspending liquid (e.g., yield-stress materials) to broaden the range of applicability of such technologies.

**Acknowledgement**

This work is carried out in the context of the project YIELDGAP (https://yieldgap-itn.eu) that has received funding from the European Union's Horizon 2020 research and innovation programme under the Marie Skłodowska Curie grant agreement No 955605.